\documentclass[authoryear,12pt,onecolumn,3p]{article}%
\usepackage{amssymb}
\usepackage{amsmath}
\usepackage{amsfonts}
\usepackage{amsbsy}
\usepackage{graphicx}
\usepackage{pifont}
\usepackage{geometry}
\usepackage[authoryear]{natbib}
\usepackage{hyperref}
\usepackage{version}
\usepackage{setspace}%
\usepackage{tabularx}
\usepackage{theorem}

\usepackage{subcaption}

\newcommand{\asterisk}{\mathord{*}}
\newcommand{\infixand}{\text{ and }}
\newcommand{\infixor}{\text{ or }}
\newcommand{\noplus}{}
\newcommand{\nospace}{}

\newcommand{\tmem}[1]{{\em #1\/}}

\newcommand{\tmmathbf}[1]{\ensuremath{\boldsymbol{#1}}}
\newcommand{\tmname}[1]{\textsc{#1}}
\newcommand{\tmrsup}[1]{\textsuperscript{#1}}
\newcommand{\tmsamp}[1]{\textsf{#1}}
\newcommand{\tmstrong}[1]{\textbf{#1}}
\newcommand{\tmtextbf}[1]{\text{{\bfseries{#1}}}}
\newcommand{\tmtextit}[1]{\text{{\itshape{#1}}}}
\newcommand{\tmtextrm}[1]{\text{{\rmfamily{#1}}}}

\newenvironment{tmparmod}[3]{\begin{list}{}{\setlength{\topsep}{0pt}\setlength{\leftmargin}{#1}\setlength{\rightmargin}{#2}\setlength{\parindent}{#3}\setlength{\listparindent}{\parindent}\setlength{\itemindent}{\parindent}\setlength{\parsep}{\parskip}} \item[]}{\end{list}}
\newenvironment{tmparsep}[1]{\begingroup\setlength{\parskip}{#1}}{\endgroup}

\newtheorem{corollary}{Corollary}
\newtheorem{lemma}{Lemma}
\newtheorem{proposition}{Proposition}
\newtheorem{theorem}{Theorem}
\newtheorem{claim}{Claim}

\providecommand{\U}[1]{\protect\rule{.1in}{.1in}}

\begin{document}
\setstretch{1.15}


\title{Coalitional Manipulations and Immunity of the Shapley Value}
\author{Christian Basteck\thanks{WZB Berlin, Berlin, Germany. Email: christian.basteck@wzb.eu.} \and Frank Huettner\thanks{SKK GSB, Sungkyunkwan University, Seoul, South Korea. Email: mail@frankhuettner.de.}\,\,\thanks{Corresponding author}}

\maketitle

\begin{abstract}
We consider manipulations in the context of coalitional games, where a coalition aims to increase the total payoff of its members. An allocation rule is immune to coalitional manipulation if no coalition can benefit from internal reallocation of worth among its subcoalitions (reallocation-proofness), and if no coalition can benefit from a lower worth (weak coalitional monotonicity). Replacing additivity in Shapley's original characterization by these requirements yields a new foundation of the Shapley value, i.e., it is the unique efficient and symmetric allocation rule that awards nothing to a null player and is immune to coalitional manipulations. We further find that for efficient allocation rules, reallocation-proofness is equivalent to constrained marginality, a weaker variant of Young's marginality axiom. Our second characterization improves upon Young’s characterization by weakening the independence requirement intrinsic to marginality.

\smallskip

 \smallskip

\end{abstract}


\section{Introduction}

In recent years, cooperative game theory and its solution methods have
expanded beyond traditional realms like cost-sharing {\citep{shubik1962,LitOwe1973,TijDri1986,gopalakrishnan2021incentives}} and
property rights remuneration {\citep{HarMoo1990,tauman2007shapley}}. Their application in statistics for identifying important variables {\citep{LipCon2001,Shorrocks2012}}, in machine learning for interpreting
prediction models {\citep{LuLe2017,Lundbergetal2020}}, and 
in marketing for attributing online advertisers' impact on customer conversion
{\citep{DalessandroEtAl2012,berman2018,Si+Be+De+Go+Iy-ManSci}} demonstrate that
today, allocation rules for cooperative games are routinely computed and
implemented {\citep{groemping2015,google2019,shap2023}}.
The most prominent allocation rule for these applications is the Shapley
value, which rewards (punishes) a player for a higher (lower) marginal contributions. In
fact, this strong monotonicity property is characteristic of the Shapley value
{\citep{young1985}}, which is commonly cited as a compelling reason for the
widespread adoption of the Shapley value {\citep{Shorrocks2012,HueSun2012,LuLe2017}}. While strong monotonicity ensures that \emph{individuals} have an incentive to work towards  a common goal, less is known about the incentives of \emph{groups of players}. 

In this paper, we study {\tmem{coalitional manipulations}}, i.e.,
modifications of a game with the intention to increase the total payoff
accruing to the members of a coalition even as the coalition does not create any additional surplus. We introduce axioms ensuring that allocation rules
are immune to such manipulations.

A most basic requirement is {\tmem{weak coalitional monotonicity}}
{\citep{zhou1991}}, which ensures that no coalition shall benefit from
having a lower worth,\footnote{Following established nomenclature, we
refer to the number $v (S)$ that is assigned to coalition $S$ in a coalitional
game as the ``worth'' of this coalition. We use the term ``value'' when
referring to a particular, well-known allocation rule, such as the Shapley
value or the equal division value. } while all else remains the same -- for example by underreporting their joint contribution to the worth of the grand coalition or by a binding pre-game agreement that reduced their worth understood as the coalition's outside option.

Inspired
by {\citet{moulin1987}}, {\citet{moulin1987}}, {\citet{thomson1988}},
{\citet{JuMiSa2007}}, and {\citet{ju2013}}, we further define
{\tmem{reallocation-proofness}} as the requirement that no coalition shall
benefit from an internal reallocation of surplus. Here, internal reallocation
means that a coalition manipulates the original game without changing (i) the
worth of the manipulating coalition itself, nor (ii) the worth of coalitions
of outside players, nor (iii) the synergies between players from the
manipulating coalition with outside players. For example a manipulating
coalition may be able to misrepresent how its subcoalitions contribute towards the worth of the grand coalition or use binding pre-game agreements on how to split up the coalitions' worth among subcoalitions should the grand coalition fail to form -- under reallocation proofness, such schemes should not improve the aggregate payoffs to the members of
the manipulating coalition.

In the context of statistics, it is not plausible that
such a manipulative initiative is the result of features or other model
entities ``becoming active''. Nonetheless, immunity to manipulation is still a
plausible requirement in the sense that it prevents a modeler or statistician
to inflate the importance of a set of features or a set of model components
through a manipulation of the game. This is particular important if the model
is otherwise noninterpretable or treated as a blackbox, and its understanding mainly relies on allocation rules from cooperative game theory like the Shapley value.

Consider for instance an XGBoost trained model that predicts a person's health
condition based on interventions features (e.g., medicine vs. placebo, special
diet plan vs. unregulated diet) and on standard individual characteristics
features (e.g., an individual's weight). Unlike a regression model, XGBoost
does not give coefficients that show how much a feature's value affects the
dependent variable. Instead, this can be done by the Shapley value, which
breaks down the contribution of each feature to a person's predicted health
condition {\citep{Lundbergetal2020}}. A prediction model might appear more
desirable to the modeler, say a pharmaceutical company, the higher the sum of
Shapley values of interventions features. Reallocation-proofness prevents that
the modeler could inflate the importance of interventions features through a
manipulation, say through different feature engineering, that merely shift
explanatory worth from one interventions features to another.

\tmtextrm{}Our main result states that the Shapley value is the unique
allocation rule that satisfies
\begin{itemize}
  \item reallocation-proofness,
  
  \item weak coalitional monotonicity,
  
  \item efficiency (the sum of payoffs equals the worth of the grand
  coalition),
  
  \item null player (a player's payoff is zero if this player's presence does
  not affect the worth of any coalition), and
  
  \item symmetry (interchangeable players get the same payoff).
\end{itemize}
The result holds true both on the domain of superadditive games, and on the
unrestricted domain of all (possibly not superadditive) cooperative games. The latter
might be appropriate, e.g., for applications in statistics. 

Notably, the
nucleolus {\citet{schmeidler1969}} satisfies all these properties except
reallocation-proofness. Hence, while the difference between the Shapley value and the
nucleolus is often pinned down to the fact that the Shapley value is monotonic
while the nucleolus is in the stable, our result offers a new perspective. We argue that the Shapley value guards against coalitions manipulating \emph{within} a game, whereas the nucleolus (and other core selectors) guards against coalitions deviating by \emph{braking away}.

If we restrict attention to efficient allocation rules, then Young's strong
monotonicity ensures immunity to manipulation. Specifically, strong
monotonicity implies weak coalitional monotonicity. Moreover, an efficient
allocation rule satisfies reallocation-proofness if and only if it satisfies
\tmtextit{constrained marginality}. The latter requires that a player's payoff
remains the same if both this player's marginal contributions {\tmem{and}} the
worth of the grand coalition remain the same. Since constrained marginality
cannot be applied if the worth of the grand coalition changes, it is a weaker
assumption than Young's marginality axiom (an undirected and weaker variant of
strong monotonicity). For example, note that the equal division value
satisfies constrained marginality but not marginality (nor strong
monotonicity).

We obtain another characterization of the Shapley value by means of
constrained marginality, weak coalitional monotonicity, efficiency, null
player, and symmetry. Our second characterization of the Shapley value sheds light on Young's
characterization by replacing strong monotonicity by the requirements of weak
coalitional monotonicity, null player (an immediate implication of strong
monotonicity given efficiency and symmetry), and constrained marginality.

According to constrained marginality, a player's payoff derives from this player's marginal contributions, but may also depend on the surplus created by other players. This notably weakens the independence requirement intrinsic to Young's marginality.
To see this more lucidly, note that constrained marginality
allows a player's payoff to depend on other players' marginal contributions in
arbitrary ways, provided that changes in other players' marginal contributions
also affect the worth of the grand coalition. Only if a player's marginal
contributions as well as the worth created by the other players remain the same,
i.e., only if a player's marginal contributions remain the same in
{\tmem{absolute}} terms and {\tmem{relative}} to the productivity of the other
players, only then does constrained marginality mandate that the player's
payoff remains the same.

The paper is organized as follows. In Section~\ref{sec.problem-formulation},
we illustrate our axioms with a basic example, and provide formal definitions.
In Section~\ref{sec.result}, we present the main result. In
Section~\ref{sec.compare.with.young}, we draw a comparison with Young's
characterization. Section~\ref{sec.disc} concludes. The appendix contains all
proofs and counterexamples, demonstrating the axioms' independence. 

\section{Problem formulation }\label{sec.problem-formulation}

We first illustrate weak coalitional monotonicity and reallocation-proofness
by a simple attribution problem in the context of advertising. Then, we
provide basic definitions and a establish a formal definition of our new
axioms.

\subsection{Example with Three Players}\label{sec:3player}

\begin{figure}[h]
    \centering
    \begin{subfigure}[b]{0.23\textwidth}
        \centering
        \includegraphics[width=\textwidth]{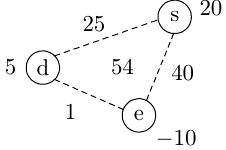}
        \caption{Original game; the Shapley value assigns the following payoffs:\\
        $
\begin{array}{lcl}
\mathrm{Sh}_d & = & 9 \\
\mathrm{Sh}_s & = & 36 \\
\mathrm{Sh}_e & = & 9 \\
\end{array}
$
     }
        \label{fig:sub1}
    \end{subfigure}
    \hfill 
    \begin{subfigure}[b]{0.23\textwidth}
        \centering
        \includegraphics[width=\textwidth]{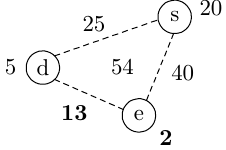}
        \caption{$\{ s, e \}$ manipulates by internal reattribution, keeping synergies with the outsider~$d$ constant:\\
        $
\begin{array}{lcl}
\mathrm{Sh}_d & = & 9 \\
\mathrm{Sh}_s & = & \mathbf{30} \\
\mathrm{Sh}_e & = & \mathbf{15} \\
\end{array}
$
  }
        \label{fig:sub2}
    \end{subfigure}
    \hfill
    \begin{subfigure}[b]{0.23\textwidth}
        \centering
        \includegraphics[width=\textwidth]{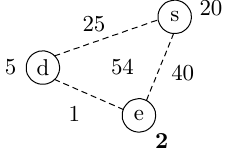}
        \caption{$\{ s, e \}$ manipulates by a reattribution that affects synergies with the outsider~$d$:\\
        $
\begin{array}{lcl}
\mathrm{Sh}_d & = & \mathbf{7} \\
\mathrm{Sh}_s & = & \mathbf{34} \\
\mathrm{Sh}_e & = & \mathbf{13} \\
\end{array}
$
    }
        \label{fig:sub3}
    \end{subfigure}
    \caption{Manipulations by coalition $\{ s, e \}$; {\textbf{changes compared to original values in (a) are bold}}; the Shapley value is immune to manipulation
by internal reattribution that do not affect synergies with outsiders, i.e.,
it satisfies reallocation-proofness, which ensures that $s$ and $e$ together
get no higher payoffs in (b) than in (a); the Shapley value is susceptible to
manipulation by reallocation that affects synergies with outsiders, i.e., it
fails strong reallocation-proofness, which would ensure that $s$ and $e$
together get no higher payoffs in (c) than in (a).}
    \label{fig:main}
\end{figure}

A company uses three advertisement services to convert customers: search ads
($s$), display ads ($d$), and email ads ($e$). The (joint) conversion scores
for the respective services are given in Figure~\ref{fig:sub1}. In the language
of cooperative games, these correspond to the worth of the various coalitions.
In this example, email by itself can be perceived as spam and has a negative
impact on customer conversion, resulting in a score of $- 10$. However, it
works well in conjunction with the other services, yielding scores of $40$ and
1 in conjunction with search and display ads, respectively. Combining all
three services amounts to a conversion score of $54.$ For simplicity, we
assume no synergies between display and search ads.

Consider the coalition of email and search (or rather their responsible
managers) and imagine that they seek to misrepresent contributions, thus
modifying the coalitional game, in order to increase their total payoff.
Suppose that the joint performance of email and search includes the positive
effect on conversion of search-triggered follow-up emails to the amount of,
say, 12. If this effect is instead ascribed to email directly, then this
yields the game of Figure~\ref{fig:sub2}. Note that for such a manipulation to
be interpreted as a reallocation among coalition members, it should have no
impact on the total worth of the manipulating coalition, i.e., on the joint
conversion score of email and search; it does however increase the joint
conversion score of email and display, as the synergies between these services
remain unchanged while additional conversions, now ascribed to email alone,
are included in conversions ascribed to display and/or email ads. Reallocation-proofness requires that such an internal reallocation within the
coalition of email and search shall not increase the total payoff that accrues
to email and search.

Next, consider the case when the manipulation by the coalition email and
search affects the synergy between email and display ads as in
Figure~\ref{fig:sub3}. In that case, a coalitional manipulation affects the
marginal contributions of the outside players (here, the marginal contribution
of $d$ to $e$ decreases by 12). Reallocation-proofness does not apply in this
case. A stronger assumption, called \tmtextit{strong reallocation-proofness},
precludes an increase in a total payoff of a manipulating coalition even when
the coalition is able to affect the marginal contributions of outsiders. The
Shapley value does not satisfy strong reallocation-proofness (in the
example, the aggregate Shapley values of email and search increase by~2). In
fact, turns out that the equal division value is the only efficient and
symmetric allocation rule that satisfies strong reallocation-proofness.

\begin{figure}[h]
    \centering
   \hspace{28mm} \begin{subfigure}[b]{0.23\textwidth}
        \centering
        \includegraphics[width=\textwidth]{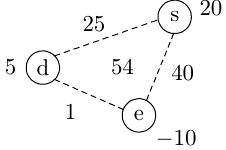}
        \caption{Original game and Shapley values:\\
        $
\begin{array}{lcl}
\mathrm{Sh}_d & = & 9 \\
\mathrm{Sh}_s & = & 36 \\
\mathrm{Sh}_e & = & 9 \\
\end{array}
$     
 }
        \label{fig:wma}
    \end{subfigure}
    \hfill 
    \begin{subfigure}[b]{0.23\textwidth}
        \centering
        \includegraphics[width=\textwidth]{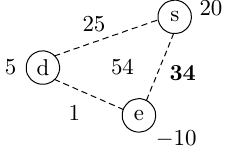}
        \caption{$\{ s, e \}$ manipulates by underreporting:\\
        $
\begin{array}{lcl}
\mathrm{Sh}_d & = & \mathbf{11} \\
\mathrm{Sh}_s & = & \mathbf{35} \\
\mathrm{Sh}_e & = & \mathbf{8} \\
\end{array}
$  
     }
        \label{fig:wmb}
    \end{subfigure}
    \hspace{28mm}
    \hfill
    \caption{Manipulation by coalition $\{ s, e \} ;$
  {\textbf{changes compared to original values in (a) are bold}}; the
  Shapley value is immune to manipulation by underreporting, i.e., it
  satisfies weak coalitional monotonicity, which ensures that $s$ and $e$
  together get no higher payoffs in (b) than in (a).}
    \label{fig:main}
\end{figure}

Finally, we deal with the situation in which the manipulating coalition is ceteris paribus 
able to lower its conversion score, say by underreporting its performance, as in Figure~\ref{fig:wmb}. Weak
coalitional monotonicity requires that this should not be advantageous to the
manipulating coalition, i.e., the total payoff to email and search shall not
increase from (a) to (b).

\subsection{Basic definitions}\label{sec.coop}

We consider coalitional games $v$: $2^N \rightarrow \mathbb{R}$, $v
(\varnothing) = 0$, where $N$ denotes the player set and~$v (S)$ the worth of
coalition $S \subseteq N$. The space of all coalitional games on $N$ is
denoted by $\bar{\mathbb{V}}$. A game $v$ is {\tmem{superadditive}} if $v (S
\cup T) \geqslant v (S) + v (T)$ for all $S, T \subseteq N$ such that $S \cap
T = \emptyset$. The collection of all superadditive coalitional games is
denoted $\mathbb{V}^s$. Moving forward, we simply refer to the domain of games
by~$\mathbb{V}$, where the results in this paper are correct if we limit all
definitions and the validity of the axioms either to the superadditive domain
(read $\mathbb{V} = \mathbb{V}^s$), or if we consider the unrestricted domain (read
$\mathbb{V} = \bar{\mathbb{V}}$).

An allocation rule $\varphi$ maps every coalitional game into payoffs, i.e.,
$\varphi$ determines for all $v \in \mathbb{V},$ and $i \in N$ a payoff
$\varphi_i (v) \in \mathbb{R}$. The {\tmem{equal division value}}
{\citet{brink2007}} assigns to each player the same share of the grand
coalition's worth,
\begin{equation}
  \begin{array}{ll}
    \mathrm{ED}_i (v) & = \frac{v (N)}{|N|} .
  \end{array} \label{eq:ed-def}
\end{equation}
The {\tmem{Shapley value}} {\citet{shapley1953}} assigns to each player its
average marginal contribution,
\begin{equation}
  \begin{array}{ll}
    \mathrm{Sh}_i (v) & = \sum_{S \subseteq N \setminus \{i\}} \frac{(|N| - 1
    - |S|) ! |S| !}{|N| !}  (v (S \cup \{i\}) - v (S)) .
  \end{array} \label{eq:sh-def}
\end{equation}
The Shapley value satisfies the following standard axioms -- and is
characterized by them.

\smallskip

\noindent\textbf{Additivity, A.}$\;$
  For all $v, w \in \mathbb{V}$, we have $\varphi  (v + w) = \varphi  (v) +
  \varphi  (w)$.\footnote{Here $(v + w) \in \mathbb{V}$ is defined by $(v + w)
  (S) = v (S) + w (S)$ for all $S \subseteq N$.}

\smallskip

\noindent\textbf{Efficiency, E.}$\;$
  For all $v \in \mathbb{V}$, we have $\sum_{i \in N} \varphi_i (v) = v (N)$.

\smallskip

\noindent\textbf{Null Player, N.}$\;$
  For all $v \in \mathbb{V}$ and all $i \in N$ we have
  \begin{eqnarray*}
    v (S \cup \{i\}) = v (S)  \hspace{1em} \mathrm{~for~all~}  S \subseteq N
    \setminus \{ i\} & \quad \Rightarrow \quad & \varphi_i (v) = 0.
  \end{eqnarray*}

\smallskip

\noindent\textbf{Symmetry, S.}$\;$
  For all $v \in \mathbb{V}$ and all $i, j \in N$ we have
  \begin{eqnarray*}
    v (S \cup \{i\})  = v (S \cup \{j\}) \hspace{1em} \mathrm{~for~all~}  S
    \subseteq N \setminus \{ i, j \} & \quad \Rightarrow \quad & \varphi_i (v)
    = \varphi_j (v) .
  \end{eqnarray*}

\begin{theorem}[{\citet{shapley1953}}]
  An allocation rule $\varphi$ satisfies additivity ({\tmstrong{A}}),
  efficiency ({\tmstrong{E}}), null player ({\tmstrong{N}}), and symmetry
  ({\tmstrong{S}}) if and only if $\varphi  \textrm{}$ is the Shapley value.
\end{theorem}

While additivity is a desirable technical property, its normative appeal is
arguably weaker than that of the other axioms. We next introduce axioms,
based on which we will provide a characterization of the Shapley value that
does not rely on additivity.

\subsection{Axioms guaranteeing immunity to manipulation}

We consider manipulations by coalitions that can be interpreted as
reallocations among its members. Hence, (i) the worth of coalitions
that consist only of outside players should be unchanged by the manipulation.
Moreover, a reallocations among its members should chancel out for the manipulating
coalition as a whole, i.e., (ii) the worth of the manipulating coalition
itself, as well as the worth of every larger coalition, should also remain
unchanged. The strongest conceivable notion of reallocation-proofness would
thus demand that the total payoffs accruing to members of a manipulating
coalition do not increase between any two games as long as (i) and (ii) are
satisfied.

\smallskip

\noindent\textbf{Strong Reallocation-proofness, }$\mathbf{R^+}$\textbf{.}$\;$
  An allocation rule $\varphi$ satisfies strong reallocation-proofness if for
  all $v, w \in \mathbb{V}$ and all $M \subseteq N$, we have
  \begin{eqnarray*}
    \left[ \begin{array}{l}
      v (T) = w (T)\\
      \mathrm{for~} \mathrm{all} \quad T \subseteq N \setminus M\\
      \mathrm{and~} \mathrm{for~} \mathrm{all} \quad T \supseteq M
    \end{array}  \right] & \Rightarrow & \sum_{i \in M} \varphi_i (v)
    \geqslant \sum_{i \in M} \varphi_i (w) .
  \end{eqnarray*}

This strong notion of reallocation-proofness turns out to be rather
restrictive. In particular, we find that the equal division value is the only
efficient and symmetric allocation rule that satisfies strong
reallocation-proofness.

\begin{proposition}
  \label{pro.s+e+r=ed} Let $| N | \neq 2.$ An allocation rule $\varphi$
  satisfies symmetry ({\tmstrong{S}}), efficiency ({\tmstrong{E}}), and strong
  reallocation-proofness ({$\mathbf{R^+}$}) if and only if $\varphi 
  \textrm{}$ is the equal division value.
\end{proposition}

Proposition~\ref{pro.s+e+r=ed}, thus begs the question how to adjust strong
reallocation-proofness in a way that preserves its intuitive interpretation of
preventing profitable reallocations. A natural variation is to further reduce
the set of feasible manipulations. Since a manipulation seems more plausible
if it remains unnoticed by the outside players, we shall limit a coalition's
manipulation to {\tmem{internal}} reallocation, i.e., reallocations that have
no impact on outside players' contributions. Conversely, where manipulations
change contributions of outside players,  reallocation-proofness is mute.

\smallskip

\noindent\textbf{Reallocation-proofness, }$\mathbf{R}$\textbf{.}$\;$
  An allocation rule $\varphi$ satisfies reallocation-proofness if for all $v,
  w \in \mathbb{V}$ and all $M \subseteq N$, we have
  \begin{eqnarray*}
    \left[ \begin{array}{l}
      v (M) = w (M) \quad \mathrm{and}\\
      v (S \cup T) - v (S)  = w (S \cup T) - w (S) \\
      \mathrm{for~} \mathrm{all~} S \subseteq M \mathrm{~and~} \mathrm{for~} \mathrm{all~} T \subseteq N \setminus M
    \end{array}  \right] & \Rightarrow & \sum_{i \in M} \varphi_i (v)
    \geqslant \sum_{i \in M} \varphi_i (w) .
  \end{eqnarray*}

Reallocation-proofness requires that the total payoff to a coalition does not
increase whenever this coalition reattributes worth internally, i.e., such
that (i) the worth of the manipulating coalition remains the same; (ii) the worth of every coalition of outside players remains the same ($S =
\emptyset$); and (iii) the surplus created by outside players remains the same.

If we limit the analysis to the domain of superadditive games,
$\mathbb{V}=\mathbb{V}^s$, then both the original game $v$ and the manipulated
game $w$ have to be superadditive. This additional domain restriction means that
reallocation-proofness only applies if the worth attributed by $M$ to its subcoalitions 
is not too large, i.e., $w (M') + w (M \setminus M') \leqslant v (M)$ for all $M'
\subseteq M$. In particular, a manipulating coalition cannot arbitrarily inflate the worth of subcoalitions or individual members.

Note that reallocation-proofness is implied by efficiency for $| N | = 2$.
Moreover, moving from strong reallocation-proofness to a weaker version
enlarges the set of reasonable allocation rules beyond the equal division
value. For example, it now includes the Shapley value.\footnote{For further
examples of rules satisfying reallocation-proofness, see \ref{sec.counterexamples}.}

\begin{lemma}
  \label{lem.sh.satisfies.R}The Shapley value satisfies reallocation-proofness
  ({\tmstrong{R}}).
\end{lemma}

Finally, we want to rule out that a coalition can benefit from a lower worth
while the game is otherwise unchanged. Formally, this is guaranteed by the
following property, introduced by {\citet{zhou1991}}.

\smallskip

\noindent\textbf{Weak Coalitional Monotonicity, }$\mathbf{W}$\textbf{.}$\;$
  An allocation rule $\varphi$ satisfies weak coalitional monotonicity if for
  all $v, w \in \mathbb{V}$ and $M \subseteq N$, we have:
  \begin{eqnarray*}
    \begin{array}{l}
    \left[  v (M)  \geqslant  w (M) \quad \mathrm{and} \quad v (S) = w (S)  \,
      \mathrm{~for~} \mathrm{all~} \, S \neq M \right]
    \end{array}  & \quad \Rightarrow \quad & \sum_{i \in M} \varphi_i (v)
    \geqslant \sum_{i \in M} \varphi_i (w) .
  \end{eqnarray*}

{\noindent}Weak coalitional monotonicity requires that the total payoff to a
coalition shall not increase whenever the worth of this coalition weakly
decreases, while the worth of all other coalitions remains the same. It is
implied by coalitional monotonicity---a property considered essential by
{\citet{shubik1962}} to prevent ``corporate idiocy'' in the context of cost
allocation.\footnote{Note that the Shapley value satisfies coalitional
monotonicity, so that weak coalitional monotonicity can be replaced by
coalitional monotonicity in our characterizations 
Theorems~\ref{thm.r+d+e+n+w=sh} and~\ref{thm.s+n+e+w+cm<less>=<gtr>Sh}.}

\smallskip

\noindent\textbf{Coalitional Monotonicity, }$\mathbf{CM}$\textbf{.}$\;$
  An allocation rule $\varphi$ satisfies coalitional monotonicity if for all
  $v, w \in \mathbb{V}$ and $M \subseteq N$, we have:
  $$
 \left[   v (M)  \geqslant  w (M) \quad \mathrm{and} \quad v (S) = w (S)  \,
    \mathrm{for~} \mathrm{all~} \, S \neq M \right] \quad \Rightarrow \quad  \varphi_i
    (v) \geqslant \varphi_i (w) .
  $$

Note, however, that weak coalitional monotonicity is a considerably weaker
assumption since its implication requires an increase only for the total
payoff of the manipulating coalition, not for each individual player. Hence,
it is not only at least as plausible and desirable, but it is further
satisfied by virtually all allocation rules considered in the literature. In
particular, weak coalitional monotonicity is compatible with core selectors
such as the nucleolus {\citep{schmeidler1969}}, which fails other monotonicity
principles such as coalitional monotonicity or monotonicity in the aggregate
{\citep{megiddo1974}}. Indeed, no core selector satisfies coalitional
monotonicity {\citep{young1985}}.\footnote{The core itself satisfies a
set-valued analogue of weak coalitional monotonicity: Consider $v, w \in \mathbb{V}$,
and $M \subseteq N$ s.t. $v (M)  \geqslant  w (M)$, and $v (S) = w (S)$ for all $\,
S \neq M$. If $\mathrm{Core} (v), \mathrm{Core} (w) \neq \emptyset$, then for every $x\in\mathrm{Core} (w)$ we have either $x\in \mathrm{Core} (v)$ or  $\sum_{i \in M} y_i > \sum_{i \in M} x_i$ for all $y\in\mathrm{Core} (v)$.}

\section{Immunity to manipulation is characteristic of the Shapley value}\label{sec.result}

Our main result states that preventing coalitional manipulation along the lines of the
axioms motivated in the previous section leaves us with a unique allocation
rule: the Shapley value.

\begin{theorem}
  \label{thm.r+d+e+n+w=sh} Let $| N | \neq 2.$ The Shapley value is the unique
  allocation rule satisfying symmetry ({\tmstrong{S}}), null player
  ({\tmstrong{N}}), efficiency ({\tmstrong{E}}), weak coalitional monotonicity
  ({\tmstrong{W}}), and reallocation-proofness ({\tmstrong{R}}).
\end{theorem}

Reallocation-proofness and weak coalitional monotonicity both rule out profitable coalitional manipulations.
Theorem~\ref{thm.r+d+e+n+w=sh} therefore provides a justification for the
Shapley value based on strategic considerations. This complements other
perspectives in cooperative game theory, in particular the question of whether
an allocation rule is stable, i.e., whether it lies in the Core {\citep{gillies1953,MoDoSh1992}}. While Core stability rules out
profitable coalitional deviations where a coalition might break away, immunity to
coalitional manipulation rules out that a coalition can remain with that grand coalition but profitable alter or misrepresent the game being played.

The proof of Theorem~\ref{thm.r+d+e+n+w=sh} is provided together with the proof of Theorem~\ref{thm.s+n+e+w+cm<less>=<gtr>Sh} in the appendix.
Notably, it is possible to prove the result both for the space of all
coalitional games as well as for the restricted domain of superadditive games.
The latter is shown in \ref{sec.supadd}. Finally, the properties in
Theorem~\ref{thm.r+d+e+n+w=sh} are independent -- counterexamples are given in
\ref{sec.counterexamples}, where we also shows that there are other allocation
rules satisfying all properties for $| N | = 2$.

\section{Comparison to Young's characterization of the Shapley value}\label{sec.compare.with.young}

A prominent characterization of the Shapley value due to {\citet{young1985}}
rests on strong monotonicity.

\smallskip

\noindent\textbf{Strong Monotonicity, }$\mathbf{M^+}$\textbf{.}$\;$
  An allocation rule $\varphi$ satisfies strong monotonicity if for all $v, w
  \in \mathbb{V}$ and all $i \in N$, we have
  \begin{eqnarray*}
    v (T \cup \{ i \}) - v (T)  \geqslant w (T \cup \{ i \}) - w (T)
    \mathrm{~for~} \mathrm{all~} T \subseteq N \setminus \{ i \} & \Rightarrow &
    \varphi_i (v) \geqslant \varphi_i (w) .
  \end{eqnarray*}

Strong monotonicity requires that a player's payoff shall not increase
whenever this player's marginal contributions to all coalitions weakly
decrease. In conjunction with efficiency and symmetry, this property is
characteristic of the Shapely value. Young emphasized that a weaker,
undirected version of strong monotonicity is sufficient to obtain uniqueness.

\smallskip

\noindent\textbf{Marginality, }$\mathbf{M}$\textbf{.}$\;$
  An allocation rule $\varphi$ satisfies marginality if for all $v, w \in
  \mathbb{V}$ and all $i \in N$, we have
  \begin{eqnarray*}
    v (T \cup \{ i \}) - v (T)  = w (T \cup \{ i \}) - w (T) \mathrm{~for~}
    \mathrm{all~} T \subseteq N \setminus \{ i \} & \Rightarrow & \varphi_i (v) =
    \varphi_i (w) .
  \end{eqnarray*}

Marginality ``is a type of independence condition rather than a monotonicity
condition'' {\citet{young1985}}, stipulating that a player's payoff shall \emph{only}
depend on this player's marginal contributions.

\begin{theorem}[{\citet{young1985}}]
  \label{thm-young-strong}An allocation rule $\varphi$ satisfies [strong
  monotonicity ({\tmstrong{M$^+$}}) or marginality ({\tmstrong{M}})],
  efficiency ({\tmstrong{E}}), and symmetry ({\tmstrong{S}}) if and only if
  $\varphi  \textrm{}$ is the Shapley value.
\end{theorem}

In view of this theorem, the Shapley value can be considered as {\tmem{the}}
allocation rule that reflects a player's merit in a game as measured by this
player's marginal contributions. \ Strong monotonicity further deters
individual players from strategically underperforming.

In contrast to strong monotonicity and marginality, weak coalitional
monotonicity carries no notion of independence, i.e., it does not require the
payoff of a player or of a coalition to be invariant with respect to all changes that do not affect their
marginal contributions. Hence, Theorems~\ref{thm.r+d+e+n+w=sh}
and~\ref{thm-young-strong} are both technically different and support
different interpretations. Young's characterization emphasizes 
individual incentives, whereas our characterization emphasizes the non-profitability of
coalitional manipulations. Note, however, that strong monotonicity implies
coalitional monotonicity {\citep{young1985}}, which in turn implies weak
coalitional monotonicity.

Even though the marginality principle has ``a long tradition in economic
theory'' {\citep{CliSer2008}}, it might be counterintuitive to require the
extend of independence that is embodied in Young's marginality axiom, i.e., that a
player's payoff cannot depend on other factors beyond this player's own marginal
contribution, and as such will be be disconnected by assumption from the wider
economic environment.\footnote{Specifically, marginality is equivalent to
requirement that a player's payoff shall only depend on this player's
dividends (see Eq. (\ref{def.div}) in the appendix for a formal definition of
dividends), i.e., marginality is equivalent to $\left[ \large{}
 d^v  (S)  = d^w  (S) \mathrm{~for~all~} S \ni
i \right] \Rightarrow \varphi_i (v) = \varphi_i (w)$. Hence, marginality rules
out that the dividends $d^v  (T), T \subseteq N \setminus \{ i \}$ can have
any influence on player $i$'s payoff.} In particular, we may be cautious to
insist on the marginality axioms in scenarios in which the worth of the grand coalition
changes dramatically. This leads us to the following weakening of marginality.

\smallskip

\noindent\textbf{Constrained Monotonicity, }$\mathbf{CM}$\textbf{.}$\;$
  An allocation rule $\varphi$ satisfies constrained marginality if for all $v, w \in
  \mathbb{V}$ and all $i \in N$, we have
  \[  \left[ \begin{array}{l}
       v (N) = w (N) \quad \mathrm{and}\\
       v (T \cup \{ i \}) - v (T)  = w (T \cup \{ i \}) - w (T) \,\mathrm{for}\,\mathrm{all}\,
     T \subseteq N \setminus \{ i \} 
     \end{array}  \right]  \,\, \Rightarrow \,\, \varphi_i (v) = \varphi_i
     (w) . \]

Similar to marginality, constrained marginality requires that a player's
payoff depends on this player's marginal contributions; however, it may also
depend on the worth of the grand coalition. Consequently, a player's payoff may increase, even if this player's {\tmem{absolute}} marginal contributions remain the same -- given that this player's marginal contributions increase
\tmtextit{}\tmtextit{relative} to the productivity of the others. Moreover,
constrained marginality is compatible with the equal division value. In
this sense, it is a clearly weaker assumption than marginality.

Interestingly, constrained marginality is closely connected to
reallocation-proofness. In fact, both principles are equivalent for rules that
satisfy efficiency.

\begin{lemma}
  \label{lem.e+r<less>=<gtr>e+cm}Let $\varphi$ satisfy efficiency
  ({\tmstrong{E}}). Then, $\varphi$ satisfies constrained marginality
  ({\tmstrong{CM}}) if and only if $\varphi$ satisfies reallocation-proofness
  ({\tmstrong{R}}).
\end{lemma}

As a consequence, if we restrict ourselves to efficient allocation rules, then
strong monotonicity implies both weak coalitional monotonicity and
reallocation-proofness. In other words, ensuring that no individual has incentives to strategically underperform
already paves the way for immunity to coalitional manipulations. Moreover, we obtain an equivalent
result to Theorem~\ref{thm.r+d+e+n+w=sh}, if we replace reallocation-proofness
by constrained marginality.

\begin{theorem}
  \label{thm.s+n+e+w+cm<less>=<gtr>Sh}Let $| N | \neq 2.$ An allocation rule
  $\varphi$ satisfies symmetry ({\tmstrong{S}}), null player ({\tmstrong{N}}),
  efficiency ({\tmstrong{E}}), weak coalitional monotonicity ({\tmstrong{W}}),
  and constrained marginality ({\tmstrong{CM}}) if and only if $\varphi =
  \mathrm{Sh}$.
\end{theorem}

\begin{figure}[h]
\centering
\includegraphics[width=0.95\textwidth]{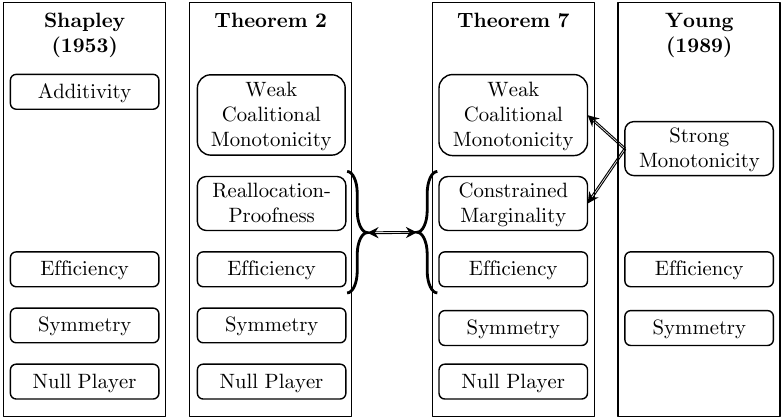}
\caption{\label{fig-results}Four characterizations of the Shapely value;
arrows indicate implications of the axioms (see
Lemma~\ref{lem.e+r<less>=<gtr>e+cm} for equivalence of
reallocation-proofness and constrained marginality in presence of efficiency;
the other implications follow from {\citet{young1985}}).}
\end{figure}

Note that all properties in Theorem~\ref{thm.s+n+e+w+cm<less>=<gtr>Sh} are
independent of each other. This indicates once more that constrained
marginality is considerably weaker than marginality. While the latter in
conjunction with symmetry and efficiency implies null player, this is not the
case for constrained marginality nor for weak coalitional monotonicity.
Moreover, since both weak coalitional monotonicity and constrained marginality
are implied by strong monotonicity, Theorem~\ref{thm.s+n+e+w+cm<less>=<gtr>Sh}
sheds light on the many consequences of the latter. Note that the null player
property is immediate from Young's axioms, so that
Theorem~\ref{thm-young-strong} is immediate from
Theorem~\ref{thm.s+n+e+w+cm<less>=<gtr>Sh} for $| N | \neq 2$.
Figure~\ref{fig-results} summarizes the mentioned Theorems and the
relationship of the utilized axioms.

Finally, we shall compare our result to {\citet{CasHue-egshmominus}}, who give
a characterization of the convex mixtures of the Shapley value and the equal
division value (the class of egalitarian Shapley values {\citep{joosten1996}})
based on weak monotonicity, which was introduced by {\citet{BrFuJu2013}}.

\smallskip

\noindent\textbf{Weak Monotonicity, }$\mathbf{M^-}$\textbf{.}$\;$
  An allocation rule $\varphi$ satisfies weak monotonicity if for all $v, w
  \in \mathbb{V}$ and all $i \in N$, we have
  \[  \left[ \begin{array}{l}
       v (N) \geqslant w (N) \quad \mathrm{and}\\
       v (T \cup \{ i \}) - v (T)  \geqslant w (T \cup \{ i \}) - w (T)
       \mathrm{~for~} \mathrm{all~} T \subseteq N \setminus \{ i \} 
     \end{array}  \right]  \quad \Rightarrow \quad \varphi_i (v) \geqslant
     \varphi_i (w) . \]

Weak monotonicity is a directed variant of constrained marginality. It is a
stronger requirement than constrained marginality and, it also implies (weak)
coalitional monotonicity. The reverse is not true, which indicates that weak
monotonicity -- despite its name -- is not an innocuous
assumption.\footnote{For example, the allocation rule given by $\varphi = 2
\mathrm{Sh} - \mathrm{ED}$ satisfies coalitional monotonicity, weak coalitional
monotonicity ({\tmstrong{W}}), constrained marginality ({\tmstrong{CM}}), and
reallocation-proofness ({\tmstrong{RP}}), but not weak monotonicity
({\tmstrong{M$^-$}}).} \ Indeed, together with symmetry and efficiency, weak
monotonicity already narrows down the set of allocation rules considerably.

\begin{theorem}[{\citet{CasHue-egshmominus}}]
  \label{thm.s+e+wm<less>=<gtr>Shalpha}Let $| N | \neq 2.$ An allocation rule
  $\varphi$ satisfies symmetry ({\tmstrong{S}}), efficiency ({\tmstrong{E}}),
  and weak monotonicity ({\tmstrong{M$^-$}}) if and only if there is an
  $\alpha \in [0, 1]$ such that $\varphi = \alpha \mathrm{Sh} + (1 - \alpha)
  \mathrm{ED}$.
\end{theorem}

Clearly, adding the null player property to
Theorem~\ref{thm.s+e+wm<less>=<gtr>Shalpha} singles out the Shapley value,
because it is the only egalitarian Shapley value that always gives nothing to
the null player.

\begin{corollary}
  \label{cor.s+e+wm+n<less>=<gtr>Sh}Let $| N | \neq 2.$ An allocation rule
  $\varphi$ satisfies symmetry ({\tmstrong{S}}), efficiency ({\tmstrong{E}}),
  weak monotonicity ({\tmstrong{M$^-$}}), and null player ({\tmstrong{N}}), if
  and only if $\varphi = \mathrm{Sh}$.
\end{corollary}

At first glance, Theorem~\ref{thm.s+n+e+w+cm<less>=<gtr>Sh} strongly resembles Corollary~\ref{cor.s+e+wm+n<less>=<gtr>Sh}, where we have the ``directionality'' in weak coalitional monotonicity. However, we remark that
we do not know of a proof that directly implies weak monotonicity from the
axioms of Theorem~\ref{thm.s+n+e+w+cm<less>=<gtr>Sh} other than the entire
uniqueness proof of Theorem~\ref{thm.s+n+e+w+cm<less>=<gtr>Sh}. In this sense,
Theorem~\ref{thm.s+n+e+w+cm<less>=<gtr>Sh} does not simply follow from
Corollary~\ref{cor.s+e+wm+n<less>=<gtr>Sh}. Moreover, it is not clear how the
proof of Theorem~\ref{thm.s+e+wm<less>=<gtr>Shalpha} by
{\citet{CasHue-egshmominus}} can be adapted to proceed within the important
subclass of superadditive games.

\section{Concluding Remarks}\label{sec.disc}

We study manipulations of cooperative games, where a coalition of players aims
to increase the total payoffs accruing to its members. Specifically, we
investigate the consequences on the payoffs assigned by the Shapley value or
alternative allocation rules. An allocation rule is immune to coalitional
manipulation if no coalition can benefit from internal reallocation of surplus
(reallocation-proofness), and if no coalition can benefit from underreporting or otherwise reducing
its worth while all else remains the same (weak coalitional monotonicity).

Replacing additivity in Shapley's original characterization by
reallocation-proofness and weak coalitional monotonicity yields a new
characterization of the value. Our characterization results are valid when
allocation rules and axioms apply either to the domain of all coalitional
games, or to the domain of superadditive games. The latter, restricted, domain not only
supports our focus on efficient allocation rules, but also makes reallocation-proofness particularly desirable:
as long as it remains within the class of superadditive games when reallocating worth among its subcoalitions, a manipulating coalition can realize the purported subcoalitions' worths by (covertly) staying together and distributing its own worth. As this renders such manipulations unobservable to outsiders, preventing them may only be achieved by ensuring that they are not in the interest of the coalition itself, i.e., by applying a rule that is reallocation-proof.

In this paper, we focus on the interpretation of coalitional manipulation by players. For applications of the Shapley value in statistics and machine learning, where features (or model components) take the role of players, immunity to manipulation is also a sound requirement. It puts limits to the extend in which a modeler or statistician can inflate the importance (measured by the Shapley value or an alternative allocation rule) of a set of features through a manipulation of the game. This aspect becomes especially crucial when the model in question is noninterpretable or treated as a blackbox, and when its assessment relies mainly on the Shapley value. Taking a model agnostic approach on the level of the cooperative game, we demonstrate that there is no general and plausible alternative allocation rule that can guarantee a higher degree of immunity to coalitional manipulation than the Shapley value. Nonetheless, the Shapley value is susceptible to coalitional manipulations if that manipulation affect the synergies of a set of features with the outside features. While this adds clarity to the question of how to manipulate the Shapley value, it also calls for further studies about manipulations of the game in specific applications.

It turns out that Young's (1985) strong monotonicity not only implies weak
coalitional monotonicity, but in conjunction with efficiency it also implies
reallocation-proofness. This allows us to shed more light on Young's
characterization based on marginality by weakening this property's notion of
independence. Marginality requires a player's
payoff to stay the same if this player's marginal contributions  stay the same, regardless of the productivity of the other players.
We show that reallocation-proofness can be replaced
by constrained marginality in our characterization. The latter is a weakening
of marginality, in the sense that it only applies if also the value of the grand
coalition remains the same, i.e., constrained marginality requires a player's
payoff to stay the same if this player's marginal contributions both absolutely and
relatively to the other players' productivity stay the same.

Reallocation-proofness and weak coalitional monotonicity ensures that an
allocation rule is immune against coalitional manipulations. This complements
other strategic considerations in cooperative game theory, in particular \ the
question of whether an allocation is stable or lies in the Core
{\citep{gillies1953,MoDoSh1992}}. While Core stability ensures
stability against coalitional {\tmem{deviation after}} allocating payoffs,
immunity to coalitional manipulation ensures stability against coalitional
{\tmem{manipulation before}} allocating payoffs.

The concept of reallocation-proofness is also helpful in view of the ``Nash
program'', i.e., the attempt to connect non-cooperative game theory and
cooperative game theory, in particular through implementations of the Shapley
value or other allocation rules via non-cooperative games (see, e.g.,
{\citet{MSPCWe2007}}, {\citet{McQSug2016}}, {\citet{BruGauMen2018}}).
Implementing an allocation rule appears more plausible if it is immune to
coalitional manipulation; moreover, we hope that our results help to connect
cooperative game theory to the study of mechanisms that rely on some concept
of coalitional strategy-proofness.

\appendix\section{Appendix}\label{app}

We first introduce further notation. Thereafter, we provide the proofs,
modifications thereof necessary to remain within the superadditive
domain{\tmsamp{}}, and counterexamples.

\subsection{Additional Notation}

For ease of notation, we denote $| N | = n$. If no confusion arises, we omit
braces around singletons. The {\tmem{marginal contribution}} of entity $i$ to
coalition $S \subseteq N \setminus i$ is denoted by $\partial_i v  (S)$,
\begin{equation}
  \partial_i v  (S) = v (S \cup i) - v (S) . \label{def.mg}
\end{equation}
We say that two players $i, j \in N$ are {\tmem{symmetric}} if $v (S \cup i) =
v (S \cup j)$ for all $S \subseteq N \setminus \{i, j\}$, i.e., if $\partial_i
v  (S) = \partial_j v  (S)$ for all $S \subseteq N \setminus \{i, j\}$. Let
$\mathbb{V}^{\ast} \subseteq \bar{\mathbb{V}}$ denote the set of symmetric
games, i.e., all players are symmetric to each other for ${{v \in
\mathbb{V}^{\ast}}^{}} $. The null game ${{\tmmathbf{0} \in
\mathbb{V}^{\ast}}^{}} $ is given by $\tmmathbf{0} (S) = 0$for all $S
\subseteq N$.

For a given $v \in \bar{\mathbb{V}}$, the dividends (also known as M{\"o}bius
inverse {\citet{rota1964}}) are recursively given by $d^v  (\varnothing) = 0$,
and
\begin{equation}
  d^v  (S) = v (S) - \sum_{R \subsetneq S} d^v  (R) \quad \text{for all } S
  \subseteq N \label{def.div} .
\end{equation}
Let $u_T$ denote the unanimity game given by $u_T (S) = 1$ if $T \subseteq S$
and otherwise $u_T (S) = 0$. It is well-known that every game has the unique
representation in unanimity games,
\begin{equation}
  v = \sum_{T \subseteq N} d^v  (T) u_T . \label{eq:decomposition}
\end{equation}
It is well-known that the Shapley value assigns to each player an equal share
of the dividends this player helps to create, i.e.,
\begin{equation}
  \mathrm{Sh}_i (v) = \sum_{S \subseteq N \text{ s.t. } i \in S} \frac{d^v 
  (S)}{| S |} . \label{eq.def.Sh.by.div}
\end{equation}
The number of non-vanishing terms in Eq.~(\ref{eq:decomposition}) is denoted
by $\#v,$ and given by
\begin{equation}
  \#v = | \{ T \subseteq N \mid d^v (T) \neq 0 \} | . \label{eq:rautev}
\end{equation}
Denote the set of players who are contained in every coalition with
non-vanishing dividend in $v$ by $R (v)$,
\begin{equation}
  R (v) = \{ i \in N \mid d^v (T) \neq 0 \Rightarrow i \in T \} .
  \label{eq:Rv}
\end{equation}
Note that $R (v) = N$ implies $d^v (T) = 0$ for $T \neq N$, i.e., $v = \lambda
u_N$ for some $\lambda \in \mathbb{R}$.

For $x \in \mathbb{R}^n$, $\bar{x} = \sum_{i \in N} x_i / n$ denotes the
average. For $x \in \mathbb{R}^n$ and $v \in \bar{\mathbb{V}}$, we define the
game $(v + x) \in \bar{\mathbb{V}}$ as the sum of $v$ and the modular game
$\sum_{i \in N} x_i {u_i} $,
\[ (v + x) (S) = v (S) + \sum_{i \in S} x_i \qquad \mathrm{for} \mathrm{all} S
   \subseteq N. \]
Finally, we write of $\varphi (x)$ instead of $\varphi (\tmmathbf{0}+ x)$.

\subsection{Proofs}

Whenever (strong) reallocation-proofness applies to a manipulation, then it
also applies to the inverse manipulation. Hence, we can imply equality of
total payoffs.

\subsubsection{Proof of Proposition \ref{pro.s+e+r=ed} on $\mathbb{\bar{V}}$}
  Consider $| N | \geq 3$ and an allocation rule $\varphi$ that satisfies
  $\tmmathbf{}$\tmtextbf{E}, \tmtextbf{S}, and
  \tmtextbf{R}\tmrsup{$\textbf{+}$}. Note that
  \tmtextbf{R}\tmrsup{$\textbf{+}$} applied to $M = N \setminus \{ i \}$ in in
  combination with \tmtextbf{E} give
  \begin{equation}
    \left[ \begin{array}{l}
      v (\{ i \}) = w (\{ i \})\\
      v (N \setminus \{ i \}) = w (N \setminus \{ i \})\\
      v (N) = w (N)
    \end{array} \right] \Rightarrow \varphi_i (v) = \varphi_i (w)
    \label{eq:grandcm}
  \end{equation}
  Towards a contradiction, assume that $\varphi$ is not the equal division
  value. Then there exists a game $v \in \bar{\mathbb{V}}$ and player $i \in
  N$ such that $\varphi_i (v) \neq v (N) / n$. Now, consider the game $w$
  constructed as follows:
  \begin{equation}
    w (S) = \left\{\begin{array}{ll}
      v (N), & S = N\\
      v (\{ i \}) + (v (N \backslash \{ i \}) - v (\{ i \})) \frac{| S | -
      1}{n - 2}, & S \subseteq N
    \end{array}\right. .
  \end{equation}
  As all players in $w$ are symmetric, $\tmmathbf{}$\tmtextbf{E} and
  $\tmmathbf{}$\tmtextbf{S} imply $\varphi_i (w) = w (N) / n = v (N) / n$.
  Yet, applying~(\ref{eq:grandcm}) gives $\varphi_i (v) = \varphi_i (w)$; a
  contradiction to $\varphi_i (v) \neq v (N) / n$.

\subsubsection{Proof of Lemma \ref{lem.sh.satisfies.R}}

Note that reallocation-proofness has an equivalent formulation referring to
dividends.

\smallskip

\noindent\textbf{Reallocation-proofness, }$\mathbf{R}$\textbf{.}$\;$
  For all $v, w \in \mathbb{V}$ and $M \subseteq N$ we have:
  \begin{eqnarray*}
    \left[ \begin{array}{l}
      \sum_{T \subseteq M} d^v  (T) = \sum_{T \subseteq M} d^w  (T) \quad
      \mathrm{and}\\
      \\
      d^v  (S) = d^v  (S) \quad \text{if } S \cap (N \setminus M) \neq
      \emptyset
    \end{array}  \right] & \quad \Rightarrow \quad & \sum_{i \in M} \varphi_i
    (v) = \sum_{i \in M} \varphi_i (w) .
  \end{eqnarray*}

With this, we can easily see that the Shapley value satisfies
reallocation-proofness.

  Using the formula $\mathrm{Sh}_i (v) = \sum_{T \subseteq N \text{ s.t. } i \in
  T} d^v (T) / | T |$, we get for all $S \subseteq N$,
  \[ \sum_{i \in S} \mathrm{Sh}_i (v) = \sum_{T \subseteq N \text{ s.t. } T
     \subseteq S} d^v  (T) + \sum_{R \subseteq N \text{ s.t. } R \cap (N
     \setminus S) \neq \emptyset} \frac{| S \cap R |}{| R |} d^v  (R) . \]

\subsubsection{Proof of Lemma \ref{lem.e+r<less>=<gtr>e+cm}}

  {\tmem{{\tmstrong{E}} and {\tmstrong{R}} imply {\tmname{{\tmstrong{CM}}}}:}}
  \ Let $v, w \in \mathbb{V}$ and $i \in N$ are such that $\partial_i v (S)  =
  \partial_i w (S) \hspace{1em} \mathrm{for} \mathrm{all} S \subseteq N \setminus i
  \large{} \hspace{1em} \mathrm{and} \hspace{1em} v (N) = w (N)$. Then, we have $v (N
  \setminus i) = w (N \setminus i)$, and applying {\tmstrong{R}} with $M = N
  \setminus i$ gives $\sum_{k \in N \setminus i} \varphi_j (v) = \sum_{k \in N
  \setminus i} \varphi_k (w)$. With {\tmstrong{E}}, we then get $\varphi_i (v)
  = \varphi_i (w)$.
  
  {\tmem{{\tmstrong{E}} and {\tmstrong{CM}} imply {\tmstrong{R}}:}} Let $v, w
  \in \mathbb{V}$ and $M \subseteq N$ are such that $\sum_{T \subseteq M} d^v 
  (T) = \sum_{T \subseteq M} d^w  (T)$ and $d^v  (S) = d^v  (S)$ if $S \cap (N
  \setminus M) \neq \emptyset$. This implies for all $j \in N \setminus M$ and
  all $S \ni j$ that $d^v  (S) = d^v  (S)$, and since for all $\hspace{1em} T
  \subseteq N \setminus j \large{}$ we have
  \begin{eqnarray*}
    \partial_j v (T)  &
    \overset{\text{(\ref{def.mg}),(\ref{eq:decomposition})}}{=} & \sum_{T'
    \subseteq (T \cup \{ i \})} d^v  (T') u_{T'} - \sum_{T' \subseteq T} d^v 
    (T') u_{T'} = \sum_{S \subseteq T : S \ni i} d^v  (S) u_{T'},
  \end{eqnarray*}
  we get $\partial_j v (T)  = \partial_j w (T) $ for all $j \in N \setminus M$
  and all $\hspace{1em} T \subseteq N \setminus j \large{}$. Since further
  \[ v (N) = \sum_{S : S \cap (N \setminus M) \neq \emptyset} d^v  (S) +
     \sum_{T \subseteq M} d^v  (T) = \sum_{S : S \cap (N \setminus M) \neq
     \emptyset} d^w  (S) + \sum_{T \subseteq M} d^w  (T) = w (N), \]
  we can apply {\tmstrong{CM}} to all players $j \in N \setminus M$ and obtain
  $\varphi_j (v) = \varphi_k (w)$ for all $j \in N \setminus M$. Finally, with
  efficiency we get
  \[ \sum_{i \in M} \varphi_i (v)  \overset{\text{{\tmstrong{E}}}}{=} v (N) -
     \sum_{j \in N \setminus M} \varphi_j (v) = w (N) - \sum_{j \in N
     \setminus M} \varphi_j (w)  \overset{\text{{\tmstrong{E}}}}{=}  \sum_{i
     \in M} \varphi_i (w), \]
  which completes the proof.

\subsubsection{Proof of Theorem \ref{thm.s+n+e+w+cm<less>=<gtr>Sh}}

It is obvious from the definition of the Shapley value (\ref{eq:sh-def}) that the above properties are satisfied by the Shapley value. Conversely, let $\varphi$ satisfy \textbf{S}, \textbf{N}, \textbf{E}, \textbf{W}, and \textbf{CM}.

\begin{claim}\label{cl.inv} For all $v \in \mathbb{V}$, $x, y \in \mathbb{R}^n$ and $i, j, k \in N$ s.t. $i, j$, and $k$ are symmetric to each other in $v$: [$\bar{x} = \bar{y}$ and $x_i = y_k$] $\Rightarrow$ $\varphi_i  (v + x ) = \varphi_k  (v + y)$.
\end{claim}

  Let $\bar{x} = \bar{y} = \mu$ and $x_i = y_k = a$. Define $x', y' \in
  \mathbb{R}^n$ as follows:
  \begin{eqnarray*}
    x'_i = a, & \quad & x'_{\ell} = \frac{n \mu - a}{n - 1} \quad \mathrm{for}
    \ell \neq i\\
    y_k' = a, & \quad & y_{\ell}' = \frac{n \mu - a}{n - 1} \quad \mathrm{for}
    \ell \neq k
  \end{eqnarray*}
  {\hspace{1.5em}}Thus, for any $j \in N \setminus \{i, k\}$, we find:
  \[ \begin{array}{ccccccll}
       \varphi_i  (v + x) & \overset{\text{{\tmstrong{CM}}\tmtextbf{}}}{=} &
       \varphi_i  (v + x') &  \overset{\text{{\tmstrong{E}}}}{=}  &  & (v +
       x') (N) - \varphi_j  (v + x') - \sum_{\ell \in N \setminus \{ i, j \}}
       \varphi_{\ell}  (v + x') &  & \\
       \varphi_k  (v + y) & \overset{\text{{\tmstrong{CM}}\tmtextbf{}}}{=} &
       \varphi_k  (v + y') &  \overset{\text{{\tmstrong{E}}}}{=}  &  & (v +
       y') (N) - \varphi_j  (v + y') - \sum_{\ell \in N \setminus \{ j, k \}}
       \varphi_{\ell}  (v + y') &  & 
     \end{array} \]
  The right hand side of both equations is the equal because: $(v + x') (N) =
  (v + y') (N)$; $\varphi_j  (v + x') = \varphi_j  (v + y')$ by
  {\tmstrong{CM}}; $\varphi_j  (v + x') = \varphi_{\ell}  (v + x')$ for all
  $\ell \in N \setminus \{ i, j \}$ by {\tmstrong{S}}; and $\varphi_j  (v +
  y') = \varphi_{\ell}  (v + y')$ for all $\ell \in N \setminus \{ j, k \}$ by
  {\tmstrong{S}}.
  
  \begin{tmparsep}{0em}%
    \begin{flushright}
      {$\diamond$} Claim \ref{cl.inv}
    \end{flushright}
  \end{tmparsep}
  
  \begin{tmparmod}{0pt}{0pt}{0pt}%
    By Claim \ref{cl.inv}, we find that for symmetric $v \in
    \mathbb{V}^{\ast}$, expressions of the form $\varphi_j  (v + z)$ depend
    only on $v, z_j$, and $\bar{z}$. This motivates the following notation:
  \end{tmparmod}
  \begin{eqnarray}
    \Delta_0^{\varphi} (v, \mu, a) & = & \varphi_i  (v + x) - \varphi_k 
    (v + y)   \label{eq.Delta}\\
    & \text{} & \mathrm{for} \hspace{1em} \mathrm{some} v \in \mathbb{V}^{\ast},
    \hspace{1em} i, k \in N, x, y \in \mathbb{R}^n \nonumber\\
    &  & \mathrm{such} \mathrm{that} \bar{x} = \bar{y} = \mu, x_i = a, y_k = 0
    \nonumber
  \end{eqnarray}
  Next, we show that $\Delta_0^{\varphi} (v, \mu, a)$ is additive in $a$ and
  therefore homogeneous \ in $a$ for rational numbers.
  
  \begin{claim}\label{cl.a.add} For all $q \in \mathbb{Q}, a , b, \mu \in \mathbb{R}$ and
    $v \in \mathbb{V}^{\ast}$: $\Delta_0^{\varphi} (v, \mu, q a + b) = q
    \Delta_0^{\varphi} (v, \mu, a) + \Delta_0^{\varphi} (v, \mu, b)$.
  \end{claim}
  
  \begin{tmparmod}{0pt}{0pt}{0pt}%
    \begin{tmparmod}{1pt}{0pt}{0pt}%
      \begin{flushleft}
        Define $x , y \in \mathbb{R}^n$ as follows:
        \[ \begin{array}{lllllllllllll}
             x_i & = & a & \qquad & x _k & = & b & \qquad & x_j & = & \frac{n
             \mu - a - b}{n - 2} & \quad & \mathrm{~for~} j \in N \setminus \{ i,
             k \}\\
             y_i & = & a + b &  & y_k & = & 0 &  & y_j & = & \frac{n \mu - a -
             b}{n - 2} &  & \mathrm{~for~} j \in N \setminus \{ i, k \}
           \end{array} \]
        Note that $\bar{x} = \mu = \bar{y}$ and $(v + x) (N) = (v + y) (N)$.
        Thus, for any $j \in N \setminus \{i, k\}$, we find
        \begin{eqnarray*}
          \Delta_0^{\varphi} (v, \mu, a + b) & \overset{\text{$\left(
          \ref{eq.Delta} \right)$}}{=} & \varphi_i  (v + y ) - \varphi_k  (v +
          y)\\
          & \overset{\text{{\tmstrong{E}}, \tmtextbf{S}}}{=} & (v + y) (N) -
          (n - 2) \varphi_j  (v + y) - \varphi_k  (v + y) - \varphi_k  (v +
          y)\\
          & \overset{\text{Claim \ref{cl.inv}} \mathrm{~for~} j}{=} & (v + y) (N)
          - (n - 2) \varphi_j  (v + x) - \varphi_k  (v + y) - \varphi_k  (v +
          y)\\
          & \overset{\text{{\tmstrong{E}}, \tmtextbf{S}}}{=} & \varphi_i  (v
          + x ) - \varphi_k  (v + y) + \varphi_k  (v + x) - \varphi_k  (v +
          y)\\
          & \overset{\text{$\left( \ref{eq.Delta} \right)$}}{=} &
          \Delta_0^{\varphi} (v, \mu, a) + \Delta_0^{\varphi} (v, \mu, b)
        \end{eqnarray*}
        Finally, it is well-known that any additive function is homogeneous in
        rational numbers.
      \end{flushleft}
    \end{tmparmod}
  \end{tmparmod}
  
  \begin{flushright}
    {$\diamond$} Claim \ref{cl.a.add}
  \end{flushright}
  
  \begin{tmparmod}{0pt}{0pt}{0pt}%
    Next, we argue that $\Delta_0^{\varphi}$ is positive if $a$ is positive
    and $v$ is the null game.
  \end{tmparmod}
  
  \begin{claim}\label{cl.a.pos} For all $a, \mu \in \mathbb{R}$: $\quad a \geqslant 0
    \quad \Rightarrow \quad \Delta_0^{\varphi}
    (\tmmathbf{0}, \mu, a) \geqslant 0$.
  \end{claim}
  
  \begin{tmparmod}{1pt}{0pt}{0pt}%
    \begin{tmparmod}{0pt}{0pt}{0pt}%
      For $a = 0$, $\Delta_0^{\varphi} (\tmmathbf{0}, \mu, a) = 0$ follows
      from definition $\left( \ref{eq.Delta} \right)$. Now let $a \neq 0$. By
      {\tmstrong{N}}, we have $\varphi_k (x) = 0$ if $x_k = 0$. Therefore,
      $\left( \ref{eq.Delta} \right)$ simplifies to
      \begin{equation}
        \Delta_0^{\varphi} (\tmmathbf{0}, \bar{x}, x_i) = \varphi_i  (x)
        \mathrm{~for~all~} x \in \mathbb{R}^n . \label{eq.deltanull}
      \end{equation}
      In particular $\Delta_0^{\varphi} (\tmmathbf{0}, \mu, \mu) = \varphi_i 
      (\mu, \ldots , \mu)$. By {\tmstrong{E}} and {\tmstrong{S}}, we
      further get $\varphi_i  (\mu, \ldots , \mu) = \mu$, i.e.,
      $\Delta_0^{\varphi} (\tmmathbf{0}, \mu, \mu) = \mu .$
      
      Now consider the case $\frac{\mu}{a} \in \mathbb{Q}$. By Claim
      \ref{cl.a.add}, $\frac{\mu}{a} \Delta_0^{\varphi} (\tmmathbf{0}, \mu, a)
      = \Delta_0^{\varphi} (\tmmathbf{0}, \mu, \mu)$. Hence,
      $\Delta_0^{\varphi} (\tmmathbf{0}, \mu, a) = \frac{a}{\mu} \mu = a,$
      i.e.,
      \begin{equation}
        \Delta_0^{\varphi} (\tmmathbf{0}, \bar{x}, x_i) = x_i \text{~if~}
        \frac{\bar{x}}{x_i} \in \mathbb{Q}. \label{eq.rational}
      \end{equation}
      Finally, let $x \in \mathbb{R}^n$ be such that $0 < x_i$. Pick some $q
      \in \mathbb{Q}$ $b$ such that $0 < q (n \bar{x} - x_i) < x_i$, and
      define $y \in \mathbb{R}^n$ by $y_i = q (n \bar{x} - x_i)$ and $y_k =
      x_k$ for $k \neq i$. Note that $\frac{\bar{y}}{y_i} = \frac{1 + q}{q}
      \in \mathbb{Q}$. By {\tmstrong{W}}, $\varphi_i  (x) > \varphi_i  (y)$.
      Hence, with (\ref{eq.deltanull}) and (\ref{eq.rational}), we get
      $\Delta_0^{\varphi} (\tmmathbf{0}, \bar{x}, x_i) > b > 0.$
      
      \begin{flushright}
        {$\diamond$} Claim \ref{cl.a.pos}
      \end{flushright}
    \end{tmparmod}
  \end{tmparmod}
  
  \begin{tmparmod}{0pt}{0pt}{0pt}%
    Next, we show that $\Delta_0^{\varphi} (\tmmathbf{0}, \mu, a)$ is
    increasing in $a$.
  \end{tmparmod}
  
  \begin{claim}\label{cl.a.mon} For all $a', a'' , \mu \in \mathbb{R}$: $a' \leqslant a''
    \Rightarrow \Delta^{\varphi} (\tmmathbf{0}, \mu, a') \leqslant
    \Delta^{\varphi} (\tmmathbf{0}, \mu, a'')$.
  \end{claim}
  
  \begin{tmparmod}{0pt}{0pt}{0pt}%
    \begin{tmparmod}{1pt}{0pt}{0pt}%
      This follows from $a'' - a' \geqslant 0$ and
      \[ \Delta^{\varphi} (\tmmathbf{0}, \mu, a'') \overset{\text{Claim
         \ref{cl.a.add} }}{=} \Delta^{\varphi} (\tmmathbf{0}, \mu, a') +
         \Delta^{\varphi} (\tmmathbf{0}, \mu, a'' - a') \overset{\text{Claim
         \ref{cl.a.pos}}}{\geqslant} \Delta^{\varphi} (\tmmathbf{0}, \mu, a')
         . \]
      \begin{flushright}
        {$\diamond$} Claim \ref{cl.a.mon}
      \end{flushright}
    \end{tmparmod}
  \end{tmparmod}
  
  \begin{tmparmod}{0pt}{0pt}{0pt}%
    Now we can argue that $\Delta_0^{\varphi} (\tmmathbf{0}, \mu, a)$ is
    linear in $a$. 
  \end{tmparmod}
  
  \begin{claim} \label{cl.a.lin} For all $\lambda, a , b, \mu \in \mathbb{R}$:
    $\Delta_0^{\varphi} (\tmmathbf{0}, \mu, \lambda a) = \lambda
    \Delta_0^{\varphi} (\tmmathbf{0}, \mu, a)$
  \end{claim}
  
  \begin{tmparmod}{0pt}{0pt}{0pt}%
    \begin{tmparmod}{1pt}{0pt}{0pt}%
      By Claim \ref{cl.a.mon}, $\Delta^{\varphi} (\tmmathbf{0}, \mu, a)$ is
      monotonic in $a$ and by Claim \ref{cl.a.add} we have $\Delta^{\varphi}
      (\tmmathbf{0}, \mu, qa) = q \Delta^{\varphi} (\tmmathbf{0}, \mu, a)$ for
      rational $q \in \mathbb{Q}.$ Since $\mathbb{Q}$ is dense in
      $\mathbb{R}$, this proves the claim.
    \end{tmparmod}
  \end{tmparmod}
  
  \begin{flushright}
    {$\diamond$} Claim \ref{cl.a.lin}
  \end{flushright}
  
  \begin{claim}   \label{cl.null} For all $x \in \mathbb{R}^n$: $\varphi (x) = \mathrm{Sh}
    (x)$.
  \end{claim}
  
  \begin{tmparmod}{1pt}{0pt}{0pt}%
    \begin{tmparmod}{0pt}{0pt}{0pt}%
      By {\tmstrong{N}}, we have $\varphi_k (\tmmathbf{0}+ x) = 0$ if $x_k =
      0$. Therefore, $\left( \ref{eq.Delta} \right)$ simplifies to
      $\Delta_0^{\varphi} (\tmmathbf{0}, \mu, \mu) = \varphi_i  (\mu, \ldots
      , \mu) .$ By {\tmstrong{E}} and {\tmstrong{S}}, we further get
      $\varphi_i  (\mu, \ldots , \mu) = \mu$, i.e., $\Delta_0^{\varphi}
      (\tmmathbf{0}, \mu, \mu) = \mu .$ By Claim \ref{cl.a.lin},
      $\Delta_0^{\varphi} (\tmmathbf{0}, \mu, a) = a$, i.e., $\varphi_i (x) -
      0 = x_i .$
      
      \begin{flushright}
        {$\diamond$} Claim \ref{cl.null}
      \end{flushright}
    \end{tmparmod}
  \end{tmparmod}
  
  Next we consider games in which cooperation requires all players.
  
  \begin{claim}
    \label{cl.un=<gtr>independent} For all $a, \mu, \lambda \in \mathbb{R}$:
    $\Delta_0^{\varphi} (\lambda u_N, \mu, a) = \Delta_0^{\varphi}
    (\tmmathbf{0}, \mu, a)$.
  \end{claim}
  
  \begin{tmparmod}{1pt}{0pt}{0pt}%
    \begin{tmparmod}{0pt}{0pt}{0pt}%
      Let $q \in \mathbb{Q}$, and define $x  \in \mathbb{R}^n$ as follows:
      \[ 
           x_i  =  q a  \qquad \qquad x _k  =  n \mu - q a \qquad \qquad  x_j  =
            0 \quad \mathrm{~for~} j \in N \setminus \{ i, k \}
        \]
      We then have
      \begin{eqnarray*}
        &  & q [\Delta_0^{\varphi} (\lambda u_N, \mu, a) - \Delta_0^{\varphi}
        (\tmmathbf{0}, \mu, a)]\\
        & \overset{\mathrm{Claim} \text{~}  \ref{cl.a.add}}{=} &
        \Delta_0^{\varphi} (\lambda u_N, \mu, q a) - \Delta_0^{\varphi}
        (\tmmathbf{0}, \mu, q a)\\
        & \overset{\text{$\left( \ref{eq.Delta} \right)$}}{=} & \varphi_i 
        (\lambda u_N + x ) - \varphi_j  (\lambda u_N + x) - [\varphi_i (x ) -
        \varphi_j (x)]\\
        & = & \varphi_i  (\lambda u_N + x ) - \varphi_i  (x ) + [\varphi_j
        (x) - \varphi_j (\lambda u_N + x)] .
      \end{eqnarray*}
      By Claim \ref{cl.inv}, $\varphi_j  (x) - \varphi_j  (\lambda u_N + x)$
      is independent of the choice of $q$. Toward a contradiction, suppose
      $\Delta_0^{\varphi} (\lambda u_N, \mu, a) \neq \Delta_0^{\varphi}
      (\tmmathbf{0}, \mu, a)$. Hence, we can find some $q$ such that
      $\varphi_i  (\lambda u_N + x ) < \varphi_i  (x )$. However, with $z$
      given by $z_i = 0$ and $z_k = \lambda / (n - 1)$,
      Claim~\ref{cl.null} implies $\varphi_i  (x ) = \varphi_i  (x  + z)$. By
      {\tmstrong{CM}}, we have $\varphi_i  (x  + z) = \varphi_i  (\lambda
      u_{N\backslash \{ i \}} + x )$. If $\lambda \geqslant 0$, then
      {\tmstrong{W}} further implies $\varphi_i  (\lambda u_{N\backslash \{ i
      \}} + x ) \leqslant \varphi_i  (\lambda u_N + x )$, i.e., \ $\varphi_i 
      (x ) \leqslant \varphi_i  (\lambda u_N + x )$ and we arrive at a
      contradiction. Analogously, for $\lambda \leqslant 0$ we can choose a
      $q$ such that $\varphi_i  (\lambda u_N + x ) > \varphi_i  (x )$ and
      construct a contradiction.
      
      \begin{flushright}
        {$\diamond$} Claim \ref{cl.un=<gtr>independent}
      \end{flushright}
    \end{tmparmod}
  \end{tmparmod}
  
  \begin{claim}
    \label{cl.un=<gtr>shapley} For all $\lambda \in \mathbb{R}$ and $x \in
    \mathbb{R}^n$: $\varphi (\lambda u_N + x) = \mathrm{Sh} (\lambda u_N + x)$.
  \end{claim}
  
  \begin{tmparmod}{1pt}{0pt}{0pt}%
    \begin{tmparmod}{0pt}{0pt}{0pt}%
      By Claims \ref{cl.un=<gtr>independent} and \ref{cl.null},
      $\Delta_0^{\varphi} (\lambda u_N, \mu, a) = \Delta_0^{\varphi}
      (\tmmathbf{0}, \mu, a) = a.$ This gives for all $x \in \mathbb{R}^n$
      \begin{eqnarray*}
        x_i - x_j  =  \Delta_0^{\varphi} (\lambda u_N, \bar{x}, x_i) -
        \Delta_0^{\varphi} (\lambda u_N, \bar{x}, x_j) = \varphi_i  (\lambda
        u_N + x ) - \varphi_j  (\lambda u_N + x) .
      \end{eqnarray*}
      Summing up over all $j \in N$ yields
      \begin{eqnarray*}
        n x_i - \sum_{j \in N} x_j & = & n \varphi_i  (\lambda u_N + x ) -
        \sum_{j \in N} \varphi_j  (\lambda u_N + x)\\
        & \overset{\text{{\tmstrong{E}}}}{=} & n \varphi_i  (\lambda u_N + x
        ) - \lambda - \sum_{j \in N} x_j .
      \end{eqnarray*}
      Hence, $\varphi_i  (\lambda u_N + x ) = \lambda / n + x_i = \mathrm{Sh}_i
      (\lambda u_N + x)$.
      
      \begin{flushright}
        {$\diamond$} Claim \ref{cl.un=<gtr>shapley}
      \end{flushright}
    \end{tmparmod}
  \end{tmparmod}
  
  \
  
  The remainder of the proof establishes $\varphi (v + x) = \mathrm{Sh} (v +
  x)$ by induction on $\#v$ as defined in~(\ref{eq:rautev}).
  
  \begin{tmparmod}{0pt}{0pt}{0pt}%
    {\tmstrong{{\tmem{{\tmem{Induction
    basis}}.}}{\tmem{}}}}$\hspace{0.27em}$For $\#v = 0$, i.e., $v =
    \textbf{0}$, $\varphi (v + x) = \mathrm{Sh} (v + x)$ follows from Claim
    \ref{cl.null}.
    
    \
    
    {\tmem{{\tmstrong{{\tmem{Induction hypothesis{\tmstrong{}}}}}}
    {\tmem{({\tmstrong{IH}}){\tmstrong{.}}}}}} Suppose $\varphi (v + x) =
    \mathrm{Sh} (v + x)$ for all $x \in \mathbb{R}^n$ and all $v \in
    \bar{\mathbb{V}}$ such that $\#v \leqslant t$.
    
    \
    
    {\tmstrong{{\tmem{{\tmem{Induction step}}}}.}} We want to show that
    \begin{equation}
      \varphi_i (v + x) = \mathrm{Sh}_i (v + x) \mathrm{~for~all~} x \in
      \mathbb{R}^n, \mathrm{~all~} i \in N, \mathrm{~and~all} v \in
      \bar{\mathbb{V}} \mathrm{~such~that}  \#v = t + 1  . \label{eq:IS}
    \end{equation}
    Let $(v + x) \in \bar{\mathbb{V}}$ be such that $\#v = t + 1$. In the case
    of $N = R (v)$, i.e., $v = \lambda u_N$ for some $\lambda \in \mathbb{R}$,
    Claim \ref{cl.un=<gtr>shapley} establishes (\ref{eq:IS}).
    
    Now consider the case $N \neq R (v)$. Pick any $i \in N \setminus R (v)$
    and construct $v^i \in \bar{\mathbb{V}}$ by
    \begin{equation}
      v^i = \sum_{T \subseteq N \text{s.t.} \nospace i \in T} d^v  (T)
      u_T . \label{eq.vi}
    \end{equation}

    Define $y  \in \mathbb{R}^n$ as follows:
    \[ 
          y_i  = x_i,  \qquad  y _k  = x_k + \sum_{T
         \subseteq N \setminus \{ i \}  \text{~s.t.~} k \in T}
         \frac{d^v  (T)}{| T |} = \mathrm{Sh}_k (v - v^i) \quad  \mathrm{~for~}
         k \in N \setminus \{ i \}
      . \]
    Note that $\#v^i \leqslant t$, $(v^i + y) (N) = (v + x) (N)$, and
    $\partial_i (v^i + y) (S) = \partial_i (v + x) (S)$ for all $S \subseteq N
    \setminus i$. Hence, we have
    \[ \varphi_i (v + x) \overset{\text{\textbf{CM}} \mathrm{~of~} \varphi}{=}
       \varphi_i (v^i + y) \overset{\text{\tmtextbf{IH}}}{=} \mathrm{Sh}_i (v^i
       + y) \overset{\text{(\ref{eq:sh-def})\tmtextbf{}}}{=} \mathrm{Sh}_i (v +
       x) . \]
    Since $i \in N \setminus R (v)$ was chosen arbitrarily, we established
    (\ref{eq:IS}) for all $i \in N \setminus R (v)$.
    
    Now pick some arbitrarily $j \in R (v)$. Choose some $k \in N \setminus R
    (v)$ and define $z  \in \mathbb{R}^n$ as follows:
    \[ \begin{array}{lllllllllllll}
         & z_k = n \bar{x} - (n - 1) x_j &  & \qquad &  &  &  &  &  &
         z_{\ell} = x_j & \quad &  & \mathrm{for} \ell \in N \setminus k
       \end{array} . \]
    Since any two player in $R (v)$ are symmetric in $(v + z)$, invoking
    {\tmstrong{S}}, {\tmstrong{E}}, and (\ref{eq:IS}) for $i \in N \setminus R
    (v)$, entails
    \begin{eqnarray*}
      | R (v) | \varphi_j (v + z) & \overset{\text{{\tmstrong{S}} of
      $\varphi$}}{=} & \sum_{\ell \in R (v)} \varphi_{\ell} (v + z)\\
      & \overset{\text{{\tmstrong{E}} of $\varphi$}}{=} & (v + z) (N) -
      \sum_{i \in N \setminus R (v)} \varphi_i (v + z)\\
      & \overset{(\ref{eq:IS}) \mathrm{~for~} i \in N \setminus R (v)}{=} & (v +
      z) (N) - \sum_{i \in N \setminus R (v)} \mathrm{Sh}_i (v + z)\\
      & \overset{ \left( \ref{eq:sh-def} \right) \text{}}{=} & | R (v) |
      \mathrm{Sh}_j (v + z) .
    \end{eqnarray*}
    This yields, $\varphi_j (v + z) = \mathrm{Sh}_j (v + z)$.
    
    Finally, since $(v + z) (N) = (v + x) (N)$, and $\partial_j (v + z) (S) =
    \partial_j (v + x) (S)$ for all $S \subseteq N \setminus \{j\}$,
    {\tmstrong{CM}} implies $\varphi_j (v + x) = \varphi_j (v + z)$ and
    $\mathrm{Sh}_j (v + x) = \mathrm{Sh}_j (v + z)$, which establishes $\varphi_j
    (v + x) = \mathrm{Sh}_j (v + x)$ for $j \in R (v)$.
  \end{tmparmod}
  
  This completes the proof on the domain $\bar{\mathbb{V}}$. The following
  section explains which adjustments are necessary for the proof to go through
  in $\mathbb{V}^s$.

\subsection{Staying within the superadditive domain
$\mathbb{V}^s$}\label{sec.supadd}

\subsubsection{Proof of Proposition \ref{pro.s+e+r=ed} on $\mathbb{V}^s$}
  To show uniqueness, suppose that $\varphi$ satisfies
  \tmtextbf{R}\tmrsup{$\textbf{+}$}, \tmtextbf{E} and \tmtextbf{S} for games
  in $\mathbb{V}^s$. Towards a contradiction, assume that there is $v \in
  \mathbb{V}^s$ and $i \in N$, s.t. $\varphi_i (v) \neq v (N) / n$. We define
  three games:
  \[ \underline{v}^1 (S) = \left\{\begin{array}{ll}
       \min   \{ v (T) : | T | = | S |, i \in T \}, & \text{if } i \in S\\
       \min \{ v (T) : | T | = | S |, i \notin T \}, & \text{if } i \notin S
     \end{array}\right. ; \]
  \[ \underline{v}^2 (S) = \left\{\begin{array}{ll}
       \min   \{ v (T) : | T | = | S | \}, & \text{if } | S | < n - 1\\
       \min \{ v (T) : | T | = | S |, i \in T \}, & \text{if } | S | = n - 1
       \infixand i \in S\\
       v (S), & \text{if } S = N \backslash \{ i \} \infixor S = N
     \end{array}\right. ; \]
  \[ \underline{v}^3 (S) = \min \{ v (T) : | T | = | S | \} . \]
  Note that the worth of the grand coalition remains $\underline{v}^1 (N) =
  \underline{v}^2 (N) = \underline{v}^3 (N) = v (N)$. All players $j \in N
  \backslash \{ i \}$ are symmetric in~$\underline{v}^1$ and
  in~$\underline{v}^2$, and all players (including player $i$) are symmetric
  in $\underline{v}^3$. Moreover, all three games are superadditive.
  
  By (8), $\varphi_i (\underline{v}^1) = \varphi_i (v) \neq v (N) / n$. Thus,
  by \tmtextbf{E} and \tmtextbf{S}, $\varphi_j (\underline{v}^1) \neq v (N) /
  n$ for all $j \in N \backslash \{ i \}$. Taking the perspective of any
  player $j \in N \backslash \{ i \}$ and applying (8) to $\underline{v}^1$
  and $\underline{v}^2$, we get $\varphi_j (\underline{v}^2) = \varphi_j
  (\underline{v}^1) \neq v (N) / n$ for all $j \in N \backslash \{ i \}$. By
  \tmtextbf{S} and \tmtextbf{E}, this implies $\varphi_i (\underline{v}^2)
  \neq v (N) / n$, i.e., $\varphi_k (\underline{v}^2) \neq v (N) / n$ for all
  $k \in N$.
  
  Finally let $\underline{k} \in \mathrm{argmin}_{k \in N} \{ v (N \backslash \{
  k \}) \}$, so that $\underline{v}^2 (N \backslash \{ \underline{k} \}) =
  \underline{v}^3 (N \backslash \{ \underline{k} \})$. Then, (8) applied to
  player $\underline{k}$, $\underline{v}^2$ and $\underline{v}^3$, gives
  $\varphi_{\underline{k}} (\underline{v}^3) = \varphi_{\underline{k}}
  (\underline{v}^2) \neq v (N) / n$. However, this is a contradiction to
  $\varphi_k (\underline{v}^3) = v (N) / n$ for all $k \in N$, which follows
  from \tmtextbf{S} and \tmtextbf{E} since all players are symmetric in
  $\underline{v}^3$.

\subsubsection{Proof of Proposition \ref{thm.s+n+e+w+cm<less>=<gtr>Sh} on $\mathbb{V}^s$}
  For all $x \in \mathbb{R}^n$, $v + x$ is superadditive if and only if $v$ is
  superadditive. The proofs of Claim~\ref{cl.inv} to
  Claim~\ref{cl.un=<gtr>shapley} stay within the class of superadditive games.
  For the remaining argument to go through, we add a second induction.
  
  First, define $\mathbb{V^{\noplus \noplus}}^+ = \{ v \in \mathbb{V} \mid
  d^v (T) \geqslant 0 \mathrm{for} \mathrm{all} T \subseteq N \} \subseteq
  \mathbb{V}^s$ as the set of games with non-negative dividends, i.e., a
  subset of all superadditive games. Note that any game derived from $v^+ \in
  \mathbb{V^{\noplus \noplus}}^+ $by deletion of dividends will remain within
  that set. Hence the induction argument in the proof of Theorem~7 on $\#v$,
  applied to $v^+ \in \mathbb{V^{\noplus \noplus}}^+$ instead of of $v \in
  \mathbb{V}$, i.e., moving to $v^+_i + y$ by deleting dividends of coalitions
  not including player~$i$ and instead distributing them equally among the
  coalitions members by an adjustment of the modular game, remains within the
  class of superadditive games. This establishes $\varphi (v  + x) = \mathrm{Sh}
  (v  + x)$ for all $v \in \mathbb{V}^+, x \in \mathbb{R}^n$.
  
  Now, take any superadditive game $v$ and define the modular game $x^v$ by
  $x^v_i = d^v (i)$ for $i \in N$. Further, we define $v^{\ast}$ and
  $\tilde{v}$ by their dividends: $d^{v^{\asterisk}} (T) = d^{\tilde{v}} (T) =
  0$ if $| T | = 1$ while for $| T | \geq 2$ we have $d^{v^{\asterisk}} (T) =
  \text{max} \{0, \text{max}_{T' : | T' | = | T |} d^v (T') \}$ and
  $d^{\tilde{v}} (T) = d^{v^{\ast}} (T) - d^v (T)$. Hence, by
  (\ref{eq:decomposition})
  \[ v^{\ast} = \sum_{T \subseteq N, | T | \geqslant 2} d^{v^{\asterisk}}
     (T) u_T  \quad \mathrm{and} \quad \tilde{v} = \sum_{T \subseteq N, | T
     | \geqslant 2}^v d^{\tilde{v}} (T) u_T \quad \mathrm{and} \quad v =
     v^{\ast} + x^v - \tilde{v} . \]
  Since ${d^{v^{\ast}} }  (T) \geqslant 0$ for all $T \subseteq N$, we have
  $v^{\ast} \in \mathbb{V}^+$ and hence in particular $\varphi (v^{\ast} +
  x^v) = \mathrm{Sh} (v^{\ast} + x^v)$.
  
  Moreover, since $d^{\tilde{v}} (T) \geq 0$ for all $T \subseteq N$, deletion
  of this dividend from $\tilde{v} $ leaves us with a superadditive game,
  i.e., $v^{\ast} + x^v - (\tilde{v} - d^{\tilde{v}} (T) u_T) = v +
  d^{\tilde{v}} (T) u_T$ is again superadditive. Thus, we can adjust the
  induction argument for a second loop as follows. The induction index is $\#
  \tilde{v}$. The induction basis becomes: For $\# \tilde{v} = 0$, i.e., $v =
  v^{\ast} + x^v$, we have $\varphi (v + x) = \mathrm{Sh} (v + x)$. The
  induction step picks $i \in N \setminus R (\tilde{v})$ and constructs $v^i
  \in \mathbb{V}$ by
  \[ v^i = v^{\ast} + x^v + \sum_{T \subseteq N \text{s.t.} \nospace i \in T}
     \widetilde{d}^v (T) u_T . \]

\subsection{Counterexamples}\label{sec.counterexamples}

The following allocation rule $\varphi^2$ satisfies all our axioms
{\tmstrong{S}}, {\tmstrong{N}}, {\tmstrong{E}}, {\tmstrong{W}},
{\tmstrong{CM}}, and {\tmstrong{R}} if there are only two players, but differs
from the Shapley value: for all $v \in \mathbb{V} (\{ i, j \})$, let

\[
\varphi_i^2 (v) = 
\begin{cases} 
    \frac{v (\{ i, j \})}{2}, & \text{if } v (i) = v (j) ;\\
    \max \{ v (i), v (\{ i, j \}) - v (j) \}, & \text{if } v (i) > v (j) ;\\
    \min \{ v (i), v (\{ i, j \}) - v (j) \}, & \text{if } v (i) < v (j) .
\end{cases}
\]

The weighted Shapley value {\citet{KalSam1987}} with unequal weights satisfies
{\tmstrong{N}}, {\tmstrong{E}}, {\tmstrong{W}}, {\tmstrong{CM}}, and
{\tmstrong{R}}, but not {\tmstrong{S}}.

The equal division value, given by $\mathrm{ED}_i (v) = v (N) / n$ for all $v
\in \mathbb{V}$ and $i \in N$, satisfies {\tmstrong{S}}, {\tmstrong{E}},
{\tmstrong{W}}, {\tmstrong{R}}, {\tmstrong{CM}}, but not {\tmstrong{N}}.

The null value, given by $\mathrm{Null}_i (v) = 0$ for all $v \in \mathbb{V}$
and $i \in N$, satisfies {\tmstrong{N}}, {\tmstrong{W}}, {\tmstrong{R}},
{\tmstrong{CM}}, and {\tmstrong{S}}, but not {\tmstrong{E}}.

The following allocation rule $\varphi^{\mathbf{W}}$ satisfies {\tmstrong{S}},
{\tmstrong{N}}, {\tmstrong{E}}, {\tmstrong{R}}, {\tmstrong{CM}}, but not
{\tmstrong{W}}:
\[ \varphi^{\mathbf{W}}_i (v) = \left\{\begin{array}{ll}
     \left( \mathrm{Sh}_i (v) - \frac{d^v (N)}{n} \right) \frac{v (N)}{v (N) -
     d^v (N)}, & \text{if } d^v (N) > 0 \infixand v (N) - d^v (N) > 0\\
     \mathrm{Sh}_i (v), & \mathrm{otherwise}
   \end{array}\right., \]
where the dividend $d^v (N)$ is defined in (\ref{def.div}).

The nucleolus satisfies {\tmstrong{S}}, {\tmstrong{N}}, {\tmstrong{E}},
{\tmstrong{W}}, but neither {\tmstrong{CM}} nor {\tmstrong{R}}.

\section*{Acknowledgments}

We thank Lars Ehlers for very helpful comments. Support by Deutsche
Forschungsgemeinschaft through CRC TRR 190 (project number 280092119) is
gratefully acknowledged.

\bibliographystyle{elsarticle-harv}\bibliography{rapshap}

\end{document}